\documentclass[a4paper,11pt]{article}

\usepackage{fullpage} 
\usepackage{parskip} 
\usepackage{tikz} 
\usepackage{amsmath}
\usepackage{hyperref}

\title{A Faster Drop-in Implementation for Leaf-wise Exact Greedy Induction of Decision Tree Using Pre-sorted Deque}
\author{Jianbo Ye \\ Pennsylvania State University \\ \texttt{jxy198@ist.psu.edu}}
\date{}

\begin{document}

\maketitle

\begin{abstract}
This short article presents a new implementation for decision trees. 
By introducing pre-sorted deques, the leaf-wise greedy tree growing strategy 
no longer needs to re-sort data at each node, and takes $O(kn)$ time and $O(1)$ extra memory locating 
the best split and branching.
The consistent, superior performance -- plus its simplicity and guarantee in producing
the same classification results as the standard
decision trees -- makes the new implementation a drop-in
replacement for leaf-wise tree induction with strong performance.
\end{abstract}

\textbf{Introduction}.
Decision tree~\cite{hunt1966experiments} is one of the oldest classification algorithms and has received increasing attentions
in recent years due to its use in building powerful ensemble models~\cite{breiman2001random,friedman2001greedy,chen2016xgboost,ke2017lightgbm}. 
It usually consists two steps:
growing a binary tree that splits data points into different leaf nodes (the induction step); prune some nodes 
from the developed tree to reduce the variation in its inferences without 
substantially degrading the classification performance (the pruning step). 
Due to its simplicity, interpretability, and generalization performance, decision tree and tree ensembles are among the most widely
adopted data analytics methods in practice.

In decision tree algorithms~\cite{quinlan1986induction,quinlan2014c4,breiman1984classification,mehta1996sliq,shafer1996sprint}, 
the induction step is often the most expensive part to compute. There have been a 
number of efforts trying to improve its speed. Some try to parallelize the tree construction process~\cite{shafer1996sprint,kufrin1997decision,jin2003communication,meng2016communication,chen2016xgboost},
some try to subsample the data points or features. A complementary approach is to speed up the algorithm itself,
which is the focus of this short article.

Prior efforts in this direction include many approximations in split-point searching, 
some of which are backed up with theoretical guarantees. They are significant contributions
but their motivations might be challenged given the new message delivered in this short article --- we will discuss this issue later. 
What's more important, the classical Hunt's algorithm~\cite{hunt1966experiments} and SPIQ~\cite{mehta1996sliq}, as well as their variants~\cite{quinlan2014c4,quinlan1986induction,breiman1984classification,shafer1996sprint}, 
still remain as the two dominant choices in practice,
exemplified by implementations such as Scikit-learn~\cite{pedregosa2011scikit}, XGBoost~\cite{chen2016xgboost}, R-package \texttt{rpart}~\cite{therneau2015rpart} and \texttt{gbm}~\cite{ridgeway2007generalized}. 
Compared to the Hunt's algorithm, SPIQ is a faster approach that pre-sorts features and grows
the tree level-wisely by scanning the entire dataset at each level. The Hunt's algorithm
was often criticized as the need to sort data points at each splitting node, and was often considered
computationally inferior to breath-wise approach SPIQ historically~\cite{mehta1996sliq,ranka1998clouds}. 
Nevertheless, Hunt's algorithm which 
adopts a leaf-wise search strategy can easily develop a very deep tree, 
a feature SPIQ would also pay a lot computations to cope with. Furthermore, the leaf-wise search
strategy realizes the possibility to prioritize which branches the tree growing process would seek
if the complete greedy tree is not constructed by the end and an early stop criterion is used instead~\cite{shi2007best}.

The message of this short article is simple and clear. We show that the leaf-wise search strategy in constructing
a decision tree is realizable with comparable or even faster performance with respect to breath-wise tree growth. For 
an unbalanced tree where most data points are sitting on only a small portion of leaf nodes, 
Hunt's algorithm can in fact be much faster than SPIQ if a proper data structure introduced in the short article is implemented. 
By maintaining pre-sorted deques, 
there is no need to sort data points at each splitting node in Hunt's algorithm. This fact in return
brings a lot speedups compared to the conventional implementation. 

\textbf{When we really need approximations}? During the past two decades, there have been great efforts in developing
approximation methods in split-point searching for a given node. There are mainly two rationales: 
\begin{enumerate}
\item \textit{High memory consumption}. 
SPIQ based approaches was claimed to be less favorable in memory performance~\cite{chen2016xgboost,ke2017lightgbm}, 
when the entire dataset is used to create node splits 
in a level-wise manner. Meanwhile, in leaf-wise tree growing --- many approximation methods adopt~\cite{jin2003communication,ranka1998clouds,ke2017lightgbm}, 
one only needs to pre-load a small portion of relevant data points to create a local branching. 
\item \textit{Inefficiency}. 
The leaf-wise greedy tree growing was often criticized by the need to re-sort features~\cite{mehta1996sliq}. Partly for the reason 
to alleviate the $O(k n \log n)$ complexity of sorting, several approximation  techniques instead search
a candidate subset of split points by grouping features into several consecutive intervals and use the interval points
as the candidates~\cite{ranka1998clouds,li2008mcrank}. Such approximation takes at least $O(k n \log m)+ O(k m)$ time at each node, where $m \ll n$ is the number of intervals, 
independent of the number of data points $n$ at the current node. 
\end{enumerate}

To be fair, the parallel improvement of SPIQ, SPRINT~\cite{shafer1996sprint} can be implemented in 
a distributed manner for large datasets with memory-resilient hash tables. 
One may wonder whether the leaf-wise tree growing strategy is more popular
because it is easier to implement. The current standard of workstation actually allows us to handle quite large dataset
to build a single decision tree with devoted memory.

If memory is not the bottleneck for pre-sorting based method, the second inefficiency argument is unlikely a strong reason
for replacing exact greedy method with approximation method to determine the best splits. Because as we will show in this short article,
leaf-wise greedy tree growing does not require re-sorting features. Once features are pre-sorted, it only takes $O(kn)$
time to find the best split points at each node. Approximation method, albeit powerful for certain areas, 
often relies on extra assumptions about the data~\cite{ranka1998clouds}, 
and can perform inferior than the standard method~\cite{chen2016xgboost} or requires extra work for 
handling special cases. For example, a quite popular practice to process missing attributes is to impute them 
with a numeric number out of the feature range. The standard exact greedy method can faithfully find the right splits to
solely analyze the effect of a missing attribute, while the approximation method has no such guarantee or require special modification
to accurately handle the missing effect.

Remind that the advancement of hardwares in the last two decades changes the picture of how we prioritize
implementing models. For an alternative algorithm to get widely accepted, we believe
that it needs to meet several requirements~\cite{ding2015yinyang}: (1) It must inherit the level of trust that 
standard algorithms (e.g. ID3~\cite{quinlan1986induction}, C4.5~\cite{quinlan2014c4}, CART~\cite{breiman1984classification}) 
has attained through the many decades of practical use; 
(2) it must produce significant speedups consistently; (3) it must be simple
to develop and deploy. 

\pagebreak 
\textbf{Our Implementation}.
Consider we have $N$ data points, and $k$ features. 
For each feature, we pre-sort the entire dataset accordingly, and save the sorted results
to a deque. Throughout the algorithm, we assume each leaf node in the tree is associated with $k$ sorted deques 
all of which are referring the same subset of data of size $n$. At the beginning, all data is associated with the root node,
and the pre-sorted deques are corresponds. 
The Hunt's decision tree algorithm proceeds with the two kinds of extra operations related to the deques. 
\begin{enumerate}
\item When the best split position is to search along a feature dimension, we traverse
 the corresponding deque by computing gains against all possible candidate splits. 
\item When a node is splitting data into two branches, the deque
corresponding this node is divided into a ``left'' deque
and a ``right'' deque. 
This is achieved by first annotating each relevant data point which branch it belongs. 
The orders of the elements in deques are preserved after splitting.
\end{enumerate}

Remark that both operations takes $O(kn)$ time, and no sorting of features is needed. The implementation details
can be found in our demo repository\footnote{\url{https://github.com/bobye/Decision_Tree_PDeque}}, where an in-memory solution is provided.
It is worth noting that the choice of using deque rather than other containers is of two folds: 
it is cache friendly in terms of traversing in the ascending order of features; it supports $O(1)$ time
inserting and removing an element at both ends such that the second operation can be achieved in linear time and constant extra memory. 

\begin{table}[htp]\centering
\begin{tabular}{rcccccccccccc}
max\_depth & 2 & 4 & 8 & 16 & 24 & 32\\\hline
\# leaf & 2 & 8 & 128 & 12,259 & 63,361 & 86,082 \\
xgboost & 0.978 & 2.119 & 4.698 & 11.565 & 19.863 & 25.469 \\
ours & 0.568 & 1.615 & 3.762 & 8.041 & 14.853 & 16.831
\end{tabular}
\caption{Seconds for constructing a single tree at different max\_depth (no node pruning). Pre-sorting time is excluded. Features are stored as 32-bit floats.}~\label{tab:result}
\end{table}

\textbf{Experiments}. We implemented Hunt's algorithm (the tree induction step) using C++11 with the aforementioned data structures.
As for performance reference, we also compare its timing to the fast implementation of SLIQ algorithm in XGBoost~\cite{chen2016xgboost} on the HIGGS Data Set\footnote{\url{https://archive.ics.uci.edu/ml/datasets/HIGGS}}. Following the practice of~\cite{chen2016xgboost}, we use a 1 million subsample to train
the decision tree. For a fair comparison, we experiment under the least conservative setting for both implementations. 
We show the performance of our serial implementation is comparable or better
than the one implemented in XGBoost (version 0.6). The goal is, however, not to demonstrate the speed superiority of Hunt's algorithm over SLIQ, but 
to show that Hunt's algorithm can be as efficient as SLIQ, which to the best of our knowledge was not known to the literature.  
The statistics are reported in Table~\ref{tab:result} based on 20 repeated runs in single-thread mode on iMac 3.3 GHz Intel Core i5.

\begin{table}[htp]\centering
\begin{tabular}{rcccccccccccc}
min\_leaf\_samples & 1 & 5 & 10 & 50 & 100 \\\hline
\# leaf & 86,082 & 31,237 & 15,146 & 3,764 & 2,205 \\
time (seconds) & 16.831 & 10.762 & 8.750 & 7.526 &  6.938
\end{tabular}
\caption{Seconds for constructing a single tree at different min\_leaf\_samples (no node pruning). Pre-sorting time is excluded.}~\label{tab:result2}
\end{table}

In the second experiment, we show the time performance of our method can be predictably improved by setting
the minimal number of samples of leaf node. We list the timing result in Table~\ref{tab:result2}. By setting
it to a larger value, the time to construct an exact greedy tree can be greatly reduced.

\textbf{Conclusion}. In this short article, we demonstrate that the Hunt's algorithm for the 
leaf-wise induction of a decision tree can be implemented
efficiently in modern computer architectures using presorted deque and 
have similar serial performance to the SLIQ~\cite{mehta1996sliq} implementation in 
highly optimized library XGBoost~\cite{chen2016xgboost}.

\scriptsize 
\bibliographystyle{plain}
\bibliography{bibliography.bib}
\end{document}